\pdfoutput=1
\documentclass[aps,prl,twocolumn,amsmath,amssymb,nofootinbib,showpacs,superscriptaddress]{revtex4}
\usepackage[english]{babel}
\usepackage{latexsym}
\usepackage{graphics}
\usepackage{epsfig}
\usepackage{color}
\usepackage{bm}
\usepackage{amsmath}
\usepackage{amssymb}
\usepackage[normalem]{ulem}
\usepackage{color}

\newcommand{\1}{\uparrow}
\newcommand{\2}{\downarrow}

\begin{document}

\title{One-dimensional three-boson problem with two- and three-body interactions}
\author{G. Guijarro}
\affiliation{Departament de F\'isica, Campus Nord B4-B5,
Universitat Polit\`ecnica de Catalunya, E-08034 Barcelona, Spain}
\author{A. Pricoupenko}
\affiliation{LPTMS, CNRS, Univ. Paris Sud, Universit\'e Paris-Saclay, 91405 Orsay, France}
\affiliation{D\'epartement de Physique, \'Ecole Normale Sup\'erieure de Lyon, 46 All\'ee d'Italie, 69007 Lyon}
\author{G.~E.~Astrakharchik}
\author{J. Boronat}
\affiliation{Departament de F\'isica, Campus Nord B4-B5,
Universitat Polit\`ecnica de Catalunya, E-08034 Barcelona, Spain}
\author{D.~S.~Petrov}
\affiliation{LPTMS, CNRS, Univ. Paris Sud, Universit\'e Paris-Saclay, 91405 Orsay, France}

%\date{\today}

\begin{abstract}

We solve the three-boson problem with contact two- and three-body interactions in one dimension and analytically calculate the ground and excited trimer-state energies. Then, by using the diffusion Monte Carlo technique we calculate the binding energy of three dimers formed in a one-dimensional Bose-Bose or Fermi-Bose mixture with attractive interspecies and repulsive intraspecies interactions. Combining these results with our three-body analytics, we extract the three-dimer scattering length close to the dimer-dimer zero crossing. In both considered cases the three-dimer interaction turns out to be repulsive. Our results constitute a concrete proposal for obtaining a one-dimensional gas with a pure three-body repulsion.

\end{abstract}

%\pacs{34.50.-s, 05.30.Jp, 67.85.-d}

\maketitle

The one-dimensional $N$-boson problem with the two-body contact interaction $g_2\delta(x)$ is exactly solvable. Lieb and Liniger \cite{LiebLiniger} have shown that for $g_2>0$ the system is in the gas phase with positive compressibility. McGuire \cite{McGuireBosons} has demonstrated that for $g_2<0$ the ground state is a soliton with the chemical potential diverging with $N$. In the case $N=\infty$ the limits $g_2\rightarrow +0$ and $g_2\rightarrow -0$ are manifestly different: The former corresponds to an ideal gas whereas the latter corresponds to collapse. Accordingly, the behavior of a realistic one- or quasi-one-dimensional system close to the two-body zero crossing strongly depends on higher-order terms not included in the Lieb-Liniger or McGuire zero-range models. Sekino and Nishida \cite{Nishida} have considered one-dimensional bosons with a pure zero-range three-body attraction and found that the ground state of the system is a droplet with the binding energy exponentially increasing with $N$, which also means collapse in the thermodynamic limit. Two of us \cite{PricoupenkoPetrov} have argued that in a sufficiently dilute regime the three-body interaction is effectively repulsive, providing a mechanical stabilization against collapse for $g_2<0$. The competition between the two-body attraction and three-body repulsion leads to a dilute liquid state similar to the one discussed by Bulgac \cite{Bulgac} in three dimensions.

The three-body scattering in one dimension is kinematically equivalent to a two-dimensional two-body scattering \cite{Petrov3body,Nishida}. Therefore, the corresponding interaction shift depends logarithmically on the product of the scattering momentum and three-body scattering length $a_3$. An important consequence of this fact is that, in contrast to higher dimensions, the one-dimensional three-body interaction can become noticeable even if $a_3$ is exponentially small compared to the mean interparticle distance. Therefore, three-body effects can be studied in the universal dilute regime essentially in any one-dimensional system that preserves a finite residual three-body interaction close to a two-body zero crossing. Universality means that the effective-range effects are exponentially small and the relevant interaction parameters are the two- and three-body scattering lengths $a_2$ and $a_3$.

In this Rapid Communication we solve the problem of three point-like bosons and analytically relate the ground and excited trimer energies with the scattering lengths. In particular, we follow the evolution of these states as the ratio $a_3/a_2$ is changed. We then consider a two-component Bose-Bose mixture with attractive interspecies and repulsive intraspecies interactions. In this system, the interspecies attraction binds atoms into dimers while the dimer-dimer interaction is tunable by changing the intraspecies repulsion \cite{PricoupenkoPetrov}. Our analytical predictions are complemented by diffusion Monte Carlo calculation of the hexamer energy permitting to determine the three-dimer scattering length close to the dimer-dimer zero crossing. We perform this procedure for equal intraspecies coupling constants and in the case where their ratio is infinite. In the latter limit one of the components is in the Tonks-Girardeau regime and the system is equivalent to a Fermi-Bose mixture. We find that the three-dimer interaction is repulsive in both cases.

Consider three bosons of mass $m$ interacting via contact two- and three-body forces characterized by the scattering lengths $a_2$ and $a_3$, respectively. The correct boundary condition for the wave function at the two-body coincidences is ensured by the two-body pseudopotential $g_2\delta(x_{ij})$ with $g_2=-2/ma_2$, where $x_{ij}=x_i-x_j$ is the distance between particles $i$ and $j$ and we set $\hbar=1$. The three-body boundary condition implies that in the limit of vanishing hyperradius $\rho=\sqrt{2/3}\sqrt{x_{12}^2+x_{13}^2+x_{23}^2}$ the three-body wave function should be proportional to $\ln(\rho/a_3)$. This small-hyperradius asymptote holds for all finite $g_2$ since at $\rho\ll |a_2|$ the two-body interaction can be neglected and the three-body kinematics corresponds to the two-dimensional scattering on a zero-range potential. The logarithmic scaling does not hold only in the case of impenetrable particles ($g_2=\infty$), where $a_3$ is ill defined. However, this case is trivial since the contact three-body interaction is completely screened by the two-body one and plays no role. The applicability conditions for the zero-range model that we use here requires, as usual, that the de Broglie wavelengths of particles be much larger than the ranges of the potentials.

In order to construct the wave function $\psi(x_1,x_2,x_3)$, let us for a moment think of it as Green's function which solves the equation
\begin{equation}\label{SchrCoord}
(\hat{H}_1+\hat{V}_2-mE)\psi(x_1,x_2,x_3)=\delta(x_{12})\delta(x_{13}),
\end{equation}
where $\hat{H}_1=-(\partial_{x_1}^2+\partial_{x_2}^2+\partial_{x_3}^2)/2$ and $\hat{V}_2=- 2[\delta(x_{12})+\delta(x_{13})+\delta(x_{23})]/a_2$. In the limit $\rho\rightarrow 0$ one can neglect $\hat{V}_2$ and $mE$ in Eq.~(\ref{SchrCoord}) which then acquires the Poisson form $-\nabla_{\boldsymbol{\rho}}^2 \psi = 2\delta(\boldsymbol{\rho})/\sqrt{3}$, where $\boldsymbol{\rho}=\{x_{12},(x_{13}+x_{23})/\sqrt{3}\}$. For small $\rho$, we thus have $\psi = -\ln(\rho/\xi)/\sqrt{3}\pi$, where $\xi$ depends on details of the full Eq.~(\ref{SchrCoord}) and is, therefore, a function of $mE$ and $a_2$. Note that if $\xi(mE,a_2)$ were equal to $a_3$, $\psi$ would satisfy the correct two- and three-body boundary conditions, thus solving our original problem. Therefore, the logic of our approach is to solve Eq.~(\ref{SchrCoord}), extract $\xi(mE,a_2)$, and find $E$ from the implicit equation $\xi(mE,a_2)=a_3$.

The solution of Eq.~(\ref{SchrCoord}) exists for any energy $E$ and is unique, if $mE$ does not belong to the spectrum of the operator $\hat{H}_1+\hat{V}_2$. Here, we will be interested in three-body bound states and will assume $E$ below the three-atom (for $a_2<0$) or atom-dimer (for $a_2>0$) scattering thresholds. Since $\hat{H}_1+\hat{V}_2$ can be diagonalized by the Bethe ansatz, one could, in principle, expand $\psi$ in terms of Bethe-ansatz states. This, however, involves the summation over a two-dimensional parameter space of free-atom states. Here we will use a different approach which allows us to work only with the trimer and atom-dimer scattering states.

Assuming zero center-of-mass momentum, we define $F(x)=2\psi(2x/3,-x/3,-x/3)/a_2$ and move $\hat{V}_2$ to the right-hand side of Eq.~(\ref{SchrCoord}) arriving at
\begin{equation}\label{PreSTM}
(\hat{H}_1-mE)\psi=\sum_{i=1}^3 F(x_i-x_j)\delta(x_{jk})+\delta(x_{12})\delta(x_{13}),
\end{equation}
where $j$ and $k$ are different from each other and from $i$. We now solve Eq.~(\ref{PreSTM}) with respect to $\psi$ by switching to momentum representation where the operator $(\hat{H}_1-mE)^{-1}$ is a number. Expressing $\psi$ in terms of $F$ and using the definition of $F$, we obtain the closed equation for $\tilde{F}(p)=\int F(x)e^{-ipx}dx$,
\begin{equation}\label{STM}
(\hat{L}-a_2/2)\tilde{F}(p)=-1/\sqrt{3p^2-4mE},
\end{equation}
where
\begin{equation}\label{L}
\hat{L}\tilde{F}(p)=\frac{\tilde{F}(p)}{\sqrt{3p^2-4mE}}+\int \frac{2\tilde{F}(q)}{p^2+pq+q^2-mE}\frac{dq}{2\pi}.
\end{equation}

The three-body contact boundary condition is taken into account by noting that $\psi$ is the sum of two functions corresponding, respectively, to the first and second terms on the right-hand side of Eq.~(\ref{PreSTM}). The former is nonsingular and equals $3\int\tilde{F}(p)(3p^2-4mE)^{-1/2}dp/2\pi$ at $\rho=0$. The latter equals $K_0(\sqrt{-mE}\rho)/\sqrt{3}\pi\approx -\ln(\sqrt{-mE}\rho e^\gamma/2)/\sqrt{3}\pi$, where $K_0$ is the decaying Bessel function and $\gamma=0.577$ is Euler's constant. The condition $\psi\propto \ln(\rho/a_3)$ then gives
\begin{equation}\label{3bodyBP}
\ln\frac{\sqrt{-mE}a_3 e^\gamma}{2}=3\sqrt{3}\pi\int \frac{\tilde{F}(q)}{\sqrt{3q^2-4mE}}\frac{dq}{2\pi}.
\end{equation}

\begin{center}
\begin{figure}[ht]
\vskip 0 pt \includegraphics[clip,width=1\columnwidth]{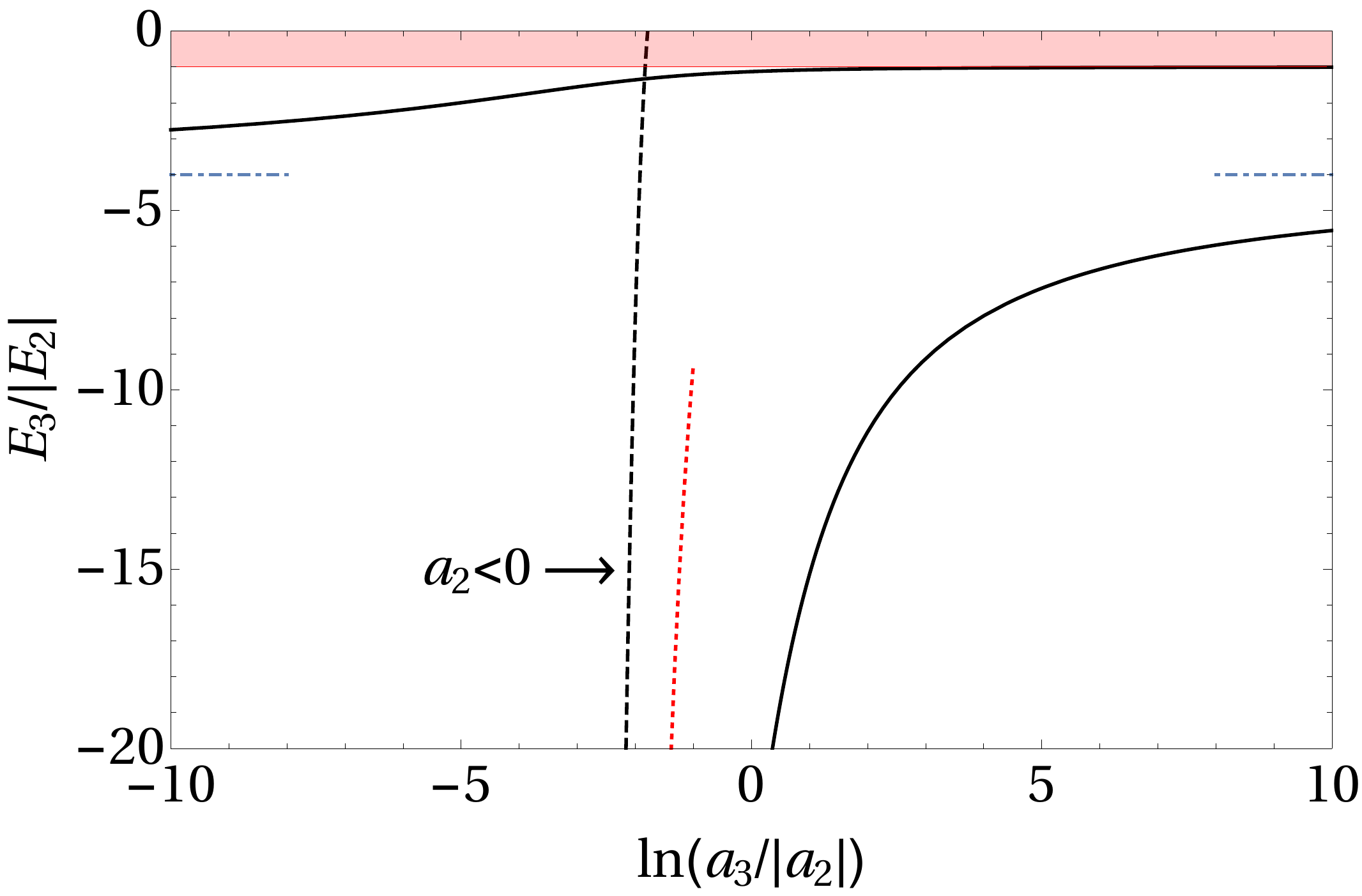}
\caption{
The trimer energy in units of $|E_2|=1/m a_2^2$ versus $\ln(a_3/|a_2|)$ for positive $a_2$ (solid black). The red filling indicates the atom-dimer scattering continuum, the blue dash-dotted lines correspond to $E_3=4E_2$ valid in the absence of the three-body force, and the red dotted line shows the asymptote $E_3=-4e^{-2\gamma}/ma_3^2$ valid in the absence of the two-body force. The black dashed curve is the trimer energy for $a_2<0$. The repulsive two-body interaction in this case pushes the trimer into the three-atom continuum at a finite value of $\ln(a_3/|a_2|)$ (see text).
}
\label{Fig:TrimerEnergy}
\end{figure}
\par\end{center}

The spectrum and eigenfunctions of $\hat{L}$ can be derived analytically from the Bethe ansatz. One can thus solve Eq.~(\ref{STM}) for $\tilde{F}$ and substitute the result into Eq.~(\ref{3bodyBP}) directly relating the trimer energy $E=E_3$ with $a_2$ and $a_3$. Although solving Eqs.~(\ref{SchrCoord}) and Eq.~(\ref{STM}) are conceptually similar tasks, the latter involves a much smaller eigenfunction basis. Note that when passing from Eq.~(\ref{SchrCoord}) to Eq.~(\ref{STM}) the roles of $E$ and $a_2$ get interchanged; $E$ is now a parameter and $a_2/2$ plays the role of an eigenvalue. Since we are dealing with $E<0$, the spectrum of $\hat{L}$ now contains only the trimer and atom-dimer scattering states. The former is characterized by the eigenfunction $\tilde{F}_{\rm McG}(p)=2(-mE)^{-1/4}/(1-p^2/mE)$ and eigenvalue $\lambda_{\rm McG}=1/\sqrt{-mE}$ consistent with the relation $E=-4/ma_2^2$ for the trimer state in the absence of three-body interaction \cite{McGuireBosons}. The continuum spectrum of $\hat{L}$ consists of atom-dimer scattering states parameterized by the atom-dimer relative momentum $k$ and characterized by eigenvalues $\lambda_k=(3k^2-4mE)^{-1/2}$. The explicit form of $\tilde{F}_k$ is obtained by Fourier transforming $F_k(x)$ extracted from the Bethe-ansatz eigenstate of $\hat{H}_1+\hat{V}_2$ with $a_2=2\lambda_k$. These manipulations result in
\begin{equation}\label{Relation}
\ln\frac{a_3\kappa e^\gamma}{a_2}=\frac{2}{\kappa^2-1}\left[\frac{\pi}{3\sqrt{3}}+\frac{3\kappa^2-1}{\sqrt{4\kappa^2-1}}\arctan\sqrt{\frac{2\kappa+1}{2\kappa-1}}\right],
\end{equation}
where $\kappa=\sqrt{-mE}a_2/2$. 

In Fig.~\ref{Fig:TrimerEnergy}, we plot $E=E_3<0$ in units of the dimer binding energy $|E_2|=1/ma_2^2$ as a function of $\ln(a_3/a_2)$ for positive $a_2$ where $E_3/E_2=4\kappa^2$. We find that there are always two trimer states in this case. For $a_3 \ll a_2$ the ground trimer is bound by the dominant three-body force and its energy tends to $-4e^{-2\gamma}/ma_3^2$ (red dotted curve). In the opposite limit $a_3\gg a_2$, the three-body interaction is subleading and the ground-trimer energy asymptotes to the McGuire result $E_3=4E_2$ \cite{McGuireBosons} (blue dot-dashed lines). The limits of large and small $a_3$ correspond to the weak three-body attraction and repulsion, respectively. The trimer follows this transition adiabatically and, in the zero-range approximation, becomes an excited state, which remains bound for any $a_3/a_2$. In the limit $a_3\rightarrow \infty$, the energy of this excited trimer asymptotically approaches the atom-dimer scattering continuum (red filled area in Fig.~\ref{Fig:TrimerEnergy}) following the threshold law $E_3/E_2-1 \approx (\pi/3)^2/\ln^2(a_3/a_2)$.

For the case $a_2<0$ (two-body repulsion), there is no dimer and $\kappa$ is negative. Equation~(\ref{Relation}) remains valid provided that its right-hand side is analytically continued from $\kappa>0$ to $\kappa<0$ just above the real axis. This gives a single trimer state, the energy of which (black dashed curve in Fig.~\ref{Fig:TrimerEnergy}) tends to $-4e^{-2\gamma}/m a_3^2$ (red dotted curve) for $a_3 \ll |a_2|$. With increasing the two-body repulsion this trimer gets pushed above the three-atom threshold at $\ln(a_3/|a_2|)=-\gamma-2\pi/3\sqrt{3}$. That we know the energy analytically makes it one of rare examples of a three-body resonance where one can study the threshold behavior to any desired order. In particular, one can show that the branch-cut singularity in this case corresponds to a two-dimensional resonance in the angular-momentum channel with $l=3$, consistent with the observation that we are dealing with a localized trimer coupled to the continuum of highly fermionized three-atom states (see the Supplemental Material of Ref.~\cite{Petrov3body}).

Returning to the two-body attraction ($a_2>0$), we note that the relative deviation of the trimer energy from the McGuire asymptote amounts to about 30\% for $a_3/a_2=e^{\pm 10}$, illustrating that even an extremely weak three-body interaction is important in one dimension. Our results can be applied to three-dimensional bosonic atoms in the quasi-one-dimensional geometry. By integrating out the radial degrees of freedom this system reduces to a pure one-dimensional model characterized by effective two- and three-body coupling constants. In the regime where the three-dimensional scattering length $a$ is much smaller than the oscillator length $l_0$ of the radial confinement, the two-body coupling constant equals $g_2 = 2a/ml_0^2$ \cite{Olshanii} and the three-body one is $g_3=-12\ln(4/3)a^2/ml_0^2$ \cite{Gora,Kovrizhin,Mazets}. On the other hand, with the logarithmic accuracy the latter can be written in terms of $a_3$ as $g_3=\sqrt{3}\pi/[m\ln(l_0/a_3)]$ \cite{PricoupenkoPetrov}. We thus identify $\ln(a_3/a_2)\approx \pi/[4\sqrt{3}\ln(4/3)]l_0^2/a^2$, which allows us to relate the trimer energies with the three-dimensional parameters $a$ and $l_0$ by using Eq.~(\ref{Relation}). Note that in this model of quasi-one-dimensional  point-like bosons the two- and three-body coupling constants vanish simultaneously with the three-dimensional scattering length $a$. Yet, three-body effects are visible and even lead to a qualitative change of the system behavior, particularly to the excited trimer state not present in the McGuire model \cite{Ludovic}.

Systems where two- and three-body effective interactions can be controlled more independently are difficult to produce or engineer (see \cite{Petrov3body} and references therein). We now discuss a model tunable to the regime of pure three-body repulsion. Namely, we consider a mixture of one-dimensional pointlike bosons $\1$ and $\2$ of unit mass characterized by the coupling constants $g_{\1\2}=-2/a_{\1\2}<0$ (interspecies attraction) and $g_{\sigma\sigma}=-2/a_{\sigma\sigma}>0$ (intraspecies repulsions). The interspecies attraction leads to the formation of $\1\2$ dimers of size $a_{\1\2}$ and energy $E_{\1\2}=-1/a_{\1\2}^2$. One can show \cite{PricoupenkoPetrov} that the two-dimer interaction changes from attractive to repulsive with increasing $g_{\sigma\sigma}$. In particular, the two-dimer zero crossing is predicted to take place for $g_{\1\1}=g_{\2\2}=2.2|g_{\1\2}|$ [Bose-Bose (BB) case] and for $g_{\2\2}=0.575|g_{\1\2}|$ if $g_{\1\1}=\infty$ [Fermi-Bose (FB) case]. Here we consider three such dimers and characterize their three-dimer interaction by calculating the hexamer energy $E_{\1\1\1\2\2\2}$ and by comparing it with the tetramer energy $E_{\1\1\2\2}$ on the attractive side of the two-dimer zero crossing where the tetramer exists. The idea is that sufficiently close to this crossing the dimers behave as pointlike particles weakly bound to each other. One can then extract the three-dimer scattering length $a_3$ from our zero-range three-boson formalism [Eq.~(\ref{Relation})] with $m=2$, $E_2=E_{\1\1\2\2}-2E_{\1\2}$, $E_3=E_{\1\1\1\2\2\2}-3E_{\1\2}$, and using the asymptotic expression for the dimer-dimer scattering length $a_2=1/\sqrt{2|E_2|}$.

In order to calculate $E_2$ and $E_3$, we resort to the diffusion Monte Carlo (DMC) technique, which is a projection method based on solving the Schr\"odinger equation in imaginary time \cite{DMC}. The importance sampling is used to reduce the statistical noise and also to impose the Bethe-Peierls boundary conditions stemming from the $\delta$-function interactions. We construct the guiding wave function $\psi_T$ in the pair-product form
\begin{equation}\label{trial wf}
\psi_T%(x_1^\1,\cdots, x_{N_\1}^\1; x_1^\2,\cdots, x_{N_\2}^\2)
=
\prod\limits_{i<j} f^{\1\1}(x_{ij}^{\1\1})
\prod\limits_{i<j} f^{\2\2}(x_{ij}^{\2\2})
\prod\limits_{i,j}f^{\1\2}(x_{ij}^{\1\2})\;,
\end{equation}
where $x_{ij}^{\sigma\sigma'} = x_i^\sigma-x_j^{\sigma'}$ is the distance between particles $i$ and $j$ of components $\sigma$ and $\sigma'$, respectively. The intercomponent correlations are governed by the dimer wave function $f^{\1\2}(x) = \exp(-|x|/a_{\1\2})$ and the intracomponent terms are $f^{\sigma\sigma}(x) = \sinh(|x|/a_{\1\2}-|x|/2a_{\rm dd})-(a_{\sigma\sigma}/a_{\1\2}-a_{\sigma\sigma}/2a_{\rm dd})$. These functions satisfy the Bethe-Peierls boundary conditions, $\partial f^{\sigma\sigma'}(x)/\partial x |_{x=+0}=-f^{\sigma\sigma'}(0)/a_{\sigma\sigma'}$, which, because of the product form, also ensures the correct behavior of the total guiding function $\psi_T$ at any two-body coincidence.
At the same time, the long-distance behavior of $f^{\sigma\sigma}(x)$ is chosen such that $\psi_T$ allows dimers to be at distances larger than their size. When the distance $x$ between pairs $\{x_1^\1, x_1^\2\}$ and $\{x_2^\1, x_2^\2\}$ is much larger than the dimer size $a_{\1\2}$, Eq.~(\ref{trial wf}) reduces to $\psi_T\propto f^{\1\2}(x_{11}^{\1\2})f^{\1\2}(x_{22}^{\1\2})\exp(-|x|/a_{\rm dd})$. For $a_{\rm dd}\gg a_{\1\2}$, this wave function describes two dimers weakly-bound to each other. While $a_{\sigma\sigma'}$ are fixed by the Hamiltonian, we treat $a_{\rm dd}$ as a free parameter in Eq.~(\ref{trial wf}). Close to the dimer-dimer zero crossing $a_{\rm dd}\approx a_2$ and this parameter is related self-consistently to the tetramer energy while far from the crossing its value is optimized according to the variational principle. It is useful to mention that in case FB, where $a_{\1\1}=0$, the $\1$ component is in the Tonks-Girardeau limit and can be mapped to ideal fermions by Girardeau's mapping \cite{Girardeau60}. Replacing $|x|$ by $x$ in the definition of $f^{\1\1}(x)$ makes $\psi_T$ antisymmetric with respect to permutations of $\1$ coordinates.

\begin{center}
\begin{figure}[ht]
\vskip 0 pt \includegraphics[clip,width=1\columnwidth]{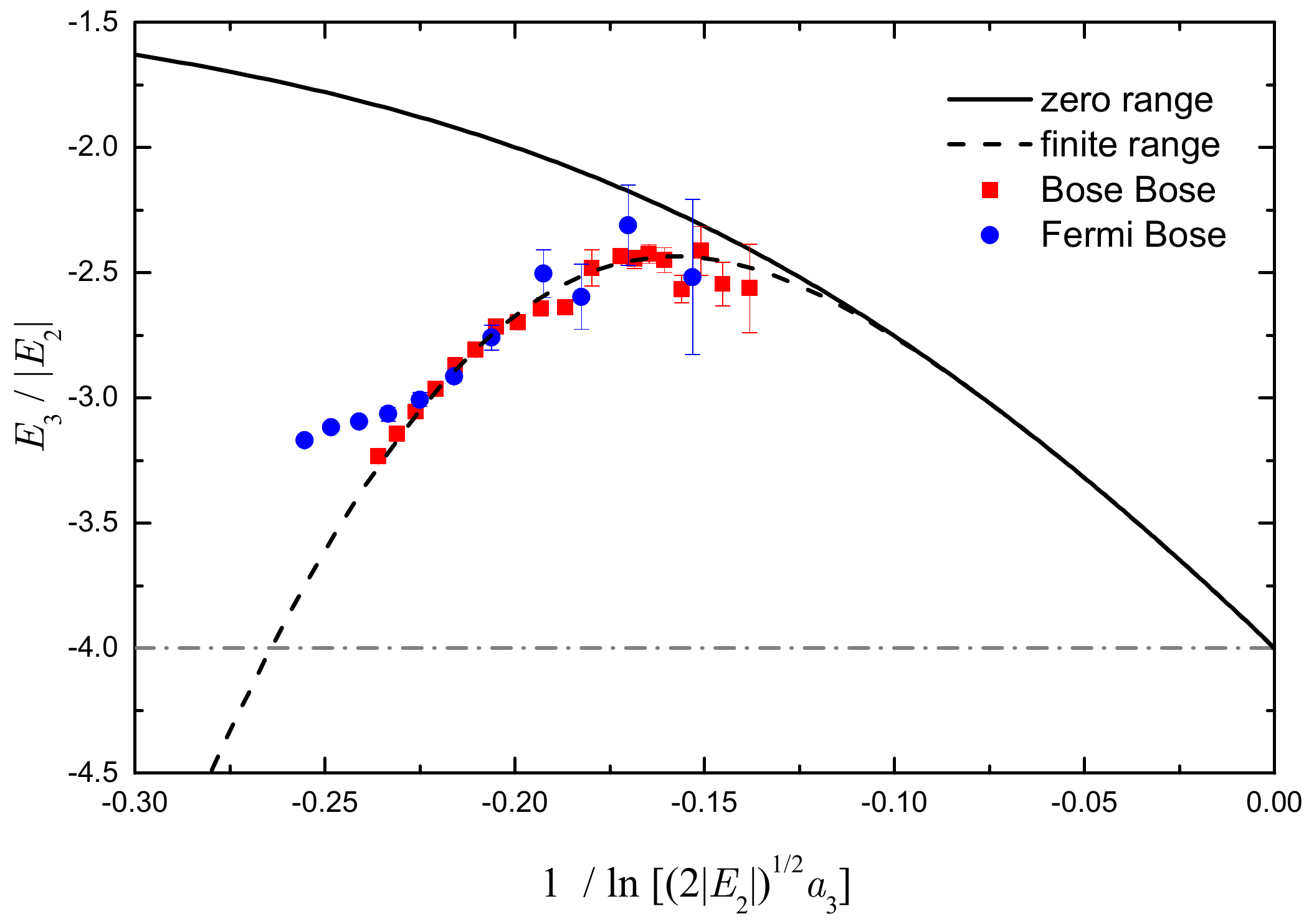}
\caption{ 
$E_3/|E_2|$ vs $1 / \ln(\sqrt{2|E_2|}a_3)$ (same as Fig.~\ref{Fig:TrimerEnergy} except for the inverse of the horizontal axis) for one-dimensional dimers. Here $E_2$ and $E_3$ are the tetramer and hexamer energies measured relative to the two- and three-dimer thresholds, respectively. The solid curve is the prediction of Eq.~(\ref{Relation}) and the dashed curve is a fit, which includes finite-dimer-size effects into account (see text). The dash-dotted line is the McGuire result $E_3=4E_2$ for three pointlike bosons with no three-body interaction. The red squares are the DMC data for case BB plotted using $a_3 = 0.01 a_{\1\2}$ and the blue circles stand for case FB with $a_3 = 0.03 a_{\1\2}$. The error bars are larger in the latter case because of the larger statistical noise induced by the nodal surface imposed by the Fermi statistics.}
\label{Fig:DMC}
\end{figure}
\par\end{center}

In Fig.~\ref{Fig:DMC}, we show $E_3/|E_2|$ for cases BB (red squares) FB (blue circles) as a function of $\delta=1/\ln(\sqrt{2|E_2|}a_3)$ along with the prediction of Eq.~(\ref{Relation}) (solid black). The quantity $a_3$ is a fitting parameter to the DMC results; changing it essentially shifts the data horizontally. We clearly see that in both cases the three-dimer interaction is repulsive since $E_3/|E_2|$ is above the McGuire trimer limit \cite{McGuireBosons} (dash-dotted line). For rightmost data points the hexamer is about ten times larger than the dimer and the data align with the universal zero-range analytics. For the other points we observe significant effective range effects related to the finite size of the dimer. In the universal limit $a_{\1\2}\ll a_2$, the leading effective-range correction to the ratio $E_3/|E_2|$ is expected to be proportional to $a_{\1\2}/a_2\propto e^{1/\delta}$ \cite{PricoupenkoPetrov}. Indeed, adding the term $C e^{1/\delta}$ to the zero-range prediction well explains deviations of our results from the universal curve and we have checked that other exponents do not work that well. We thus treat $a_3$ and $C$ as fitting parameters; in case BB we obtain $a_3=0.01a_{\1\2}$ and in case FB $a_3=0.03 a_{\1\2}$. Both cases are fit with $C=-100$ (dashed curve in Fig.~\ref{Fig:DMC}). We emphasize that we are dealing with the true ground state of three dimers. The lower ``attractive'' state formally existing for these values of $a_2$ and $a_3$ in the zero-range model is an artifact since it does not satisfy the zero-range applicability condition. The three-dimer interaction is an effective finite-range repulsion which supports no bound states.

In conclusion, we obtain an analytical expression for the ground and excited trimer energies for one-dimensional bosons interacting via zero-range two- and three-body forces. We argue that since in one dimension the three-body energy correction scales logarithmically with the three-body scattering length $a_3$, three-body effects are observable even for exponentially small $a_3$, which significantly simplifies the task of engineering three-body-interacting systems in one dimension. We demonstrate that Bose-Bose or Fermi-Bose dimers, previously shown to be tunable to the dimer-dimer zero crossing, exhibit a noticeable three-dimer repulsion. We can now be certain that the ground state of many such dimers slightly below the dimer-dimer zero crossing is a liquid in which the two-body attraction is compensated by the three-body repulsion \cite{Bulgac,PricoupenkoPetrov}.

Our results have implications for quasi-one-dimensional mixtures. We mention particularly the $^{40}$K-$^{41}$K Fermi-Bose mixture which emerges as a suitable candidate for exploring the liquid state of fermionic dimers. Here the intraspecies $^{41}$K-$^{41}$K background interaction is weakly repulsive (the triplet $^{41}$K-$^{41}$K scattering length equals 3.2nm \cite{Falke}) and the interspecies one features a wide Feshbach resonance at 540G \cite{Zwierlein}. Let us identify $\1$ with $^{40}$K, $\2$ with $^{41}$K, and assume the radial oscillator length $l_0=56$nm, which corresponds to the confinement frequency $2\pi\times 80$kHz. Under these conditions the effective coupling constants equal $g_{\sigma\sigma'}\approx 2a_{\sigma\sigma'}^{\rm (3D)}/l_0^2$ \cite{Olshanii} and the dimer-dimer zero crossing at $g_{\2\2}=0.575|g_{\1\2}|$ is realized for the three-dimensional scattering lengths $a_{\2\2}^{\rm (3D)}\approx 3.2$nm and $a_{\1\2}^{\rm (3D)}\approx-5.6$nm. The dimer size is then $\approx 560$nm and dimer binding energy corresponds to $\approx 2\pi \times 800$Hz placing the system in the one-dimensional regime. For the rightmost (next to rightmost) blue circle in Fig.~\ref{Fig:DMC}, the tetramer is approximately 20 (10) times larger than the dimer and 800 (200) times less bound. Moving left in this figure is realized by increasing $|a_{\1\2}^{\rm 3D}|$ and thus getting deeper in the region $g_{\2\2}<0.575|g_{\1\2}|$. Note, however, that this also pushes the system out of the one-dimensional regime and effects of transversal modes \cite{Gora,Kovrizhin,Mazets} become important.

While completing this paper, we became aware of a related work \cite{NishidaB} reporting the solution of the three-boson problem with zero-range interactions.

The research leading to these results received funding from the European Research Council (FR7/2007-2013 Grant Agreement No. 341197) and the MICINN (Spain) Grant No. FIS2014-56257-C2-1-P. We acknowledge the computer resources at MareNostrum and the technical support provided by Barcelona Supercomputing Center (FI-2017-3-0023). G.G. acknowledges a fellowship by CONACYT (Mexico).

G.G. and A.P. contributed equally to this work.

\end{document}